# Intelligent Receivers for Electronic Warfare Applications

A. K. Karthik[1], Jameer Ali M.S[2], A. Bhagavathi Rao[3]
UURMI systems Pvt. Ltd., Door No: 860/A, Road No: 45,
Jubilee Hills, Hyderabad, A.P., India
[1]krishnakarthika@uurmi.com
[2]jameeralim@uurmi.com
[3]bhagavatrao123@gmail.com

*Abstract*:
*In this paper, we propose an algorithm to perform modulation classification on a 5–class problem consisting of {AM, 2-PSK, 4-PSK, 8-PSK, 16-QAM} modulation schemes using a combination of features based on the first order cyclostationarity, second and higher-order moments, and then extend the idea of classification to an intelligent receiver which classifies and demodulates the signal without prior information regarding the transmitted signal.*

*Keywords— Classification, Higher-order statistics, moments, fading, adaptive receivers*

## I. INTRODUCTION

Electronic warfare (EW) refers to any action involving the use of the electromagnetic spectrum or directed energy to control the spectrum, attack an enemy, or impede enemy assaults via the spectrum. The purpose of electronic warfare is to deny the opponent the advantage of, and ensure friendly unimpeded access to, the electromagnetic spectrum. Sensing, classifying, and adapting signals are an important aspect of EW and Cognitive Radio.

Modulation Classification refers to the process of identifying the modulation format of the transmitted signal from the received signal. It is the intermediate operation between detection and demodulation, and plays an important role in military applications such as surveillance and electronic warfare. In the presence of various practical problems such as carrier frequency offset, carrier phase offset, timing offset and multipath fading channel, modulation classification becomes a challenging task [1]. The approaches followed in modulation classification can be broadly divided into two groups: decision-theoretic approach [2] and feature-based approach [3]-[7], [11]. The decision-theoretic classifiers are optimal, but they generally suffer from high computational complexity, which makes them practically not feasible. On the other hand, feature-based methods rely on the features derived from the data for modulation classification. A library of features used for classification is usually derived off-line, and decision is made based on the best match of the features estimated from real-time finite data with those in the library. The commonly used features are based on higher-order statistics such as cumulants, moments and cyclic cumulants.

In contrast to the decision-theoretic methods, the feature-based methods are non-optimal, but they are simple to implement and can often yield performance close to optimal, if carefully designed. Although there are many algorithms available in the literature for modulation classification, very few of them are practically feasible for real time systems due to their high computational complexity.

In this paper, we propose and evaluate an algorithm for performing classification on a 5–class problem consisting of the modulation schemes: {AM, 2-PSK, 4-PSK, 8-PSK, and 16-QAM}, and then extend the idea of classification to an intelligent receiver which can classify as well as demodulate {AM, 2-PSK} signals without *a priori* information regarding the transmitted signal.

In our approach, we use the concept of first-order cyclostationarity to separate AM from PSK, QAM modulations by determining the number of first-order cycle frequencies. We then find the sample, in the oversampled received signal, which is closest to the symbol time and obtain the symbol spaced sequence. We derive suitable features from the symbol spaced sequence and perform classification among the PSK and QAM modulation schemes with different constellation sizes using a combination of features based on second-, and fourth-order moments. The proposed algorithm has been designed for flat fading scenarios, but can be easily extended to other more challenging scenarios such as multipath, similar to our work in [11]. In the proposed intelligent receiver, we use the algorithm described above to perform classification, and then demodulate and extract the information from the received samples without *a priori* information regarding the training sequence. The proposed algorithms have been implemented on a USRP COTS platform, and the algorithm work in real-time scenarios performing classification as well demodulation indicating the practical feasibility of the proposed algorithms.

In Section II, we present the signal model along with the considered problem statement. In Section III, we present the used approach for timing synchronization. In Section IV, we present some information regarding first order cyclostationarity. In Section V, we present the various features used in our blind modulation classification algorithm. In Section VI & VII, we present the algorithms developed for modulation classification and the intelligent receiver. In Section VIII, we present some simulation results highlighting the performance of the proposed classification algorithm. In Section IX, we present some information regarding the implementation of these algorithms in the COTS platform, and finally in Section X, we present the conclusion of the paper.

## II. SIGNAL MODEL AND PROBLEM STATEMENT

## 1. Signal Model

The received signal, $y(t)$, received in the presence of carrier frequency/phase offsets, flat-fading channel and noise is modelled as

$$y(t) = c_0 e^{j2\pi f_o t + j\theta} x(t) + b(t) \quad (1)$$

Where $x(t)$, represents the transmitted signal, $c_0$ represents the complex channel gain, $f_o$ represents the frequency offset, $\theta$ represents the phase offset, $b(t)$ represents the additive white Gaussian noise of variance $\sigma^2$. The received discrete time signal $y(iT_s)$ can be modelled as

$$y(iT_s) = c_0 e^{j2\pi f_o iT_s + j\theta} x(iT_s - \epsilon T) + b(iT_s) \quad (2)$$

Where $T_s$ represents the sampling time, $\epsilon$ represents the timing offset, and $T$ represents the symbol time. The transmitted signal in the case of 2-PSK, 4-PSK, 8-PSK, and 16-QAM is modelled as

$$x(t) = \sum_{k=-\infty}^{\infty} s_k g(t - kT) \quad (3)$$

Where $s_k$ represents the symbol from digital modulation, $g(t)$ represents the pulse shape used which generally has a raised cosine spectrum with a roll-off $\rho$. We assume the symbols are zero mean and uncorrelated i.e., $E[s_k] = 0$, $E[s_k s_m^*] = \delta_{km}$, where $\delta_{km}$ represents the Kronecker delta function. The transmitted signal in the case of AM is modelled as

$$x(t) = (1 + K_a m(t)) \quad (4)$$

Where $m(t)$ represents the transmitted analog signal, and $K_a$ represents the modulation index of AM.

## 2. Problem Statement

In this paper, we address the problem of modulation classification for a 5-class problem consisting of the following modulation formats: {AM, 2-PSK, 4-PSK, 8-PSK, 16-QAM} in the presence of carrier frequency/phase, and timing offsets and flat-fading scenarios. We assume that an estimate of the noise variance $\sigma^2$ is available a priori. We also extend the idea of classification to design an intelligent receiver which can classify as well demodulate AM and 2-PSK modulation schemes from the received samples without any *prior* information of the training sequence.

## III. ESTIMATION OF THE SAMPLE CLOSEST TO THE SYMBOL TIME

The method is based on [9]. The discrete-time version of the received signal in the case of M-PSK and M-QAM after the match filtering operation with a root-raised cosine pulse shape $h_{rrc}(.)$ of roll-off $\rho$ is given by

$$r(iT_s) = c_0 e^{j2\pi f_o iT_s + j\theta} \sum_{k=-\infty}^{\infty} s_k h(iT_s - kT - \epsilon T) + n(iT_s) \quad (5)$$

Where $h(.)$ represents the overall pulse shape, and $n(.) = b(.) \otimes h_r(.)$ represents the filtered noise. As proposed in [9], the estimate of timing offset $\hat{\epsilon}$ is given by

$$\hat{\epsilon} = \frac{-1}{2\pi} arg(X_m) \quad (6)$$

Where $X_m$ is given by

$$X_m = \sum_{i=0}^{P\mathcal{N}-1} |r(iT_s)|^2 e^{-j2\pi i/P} \quad (7)$$

Where $P$ represents the oversampling factor and $\mathcal{N}$ represents the number of symbols used for estimation. We then choose the sample closest to the symbol timing (sample having the smallest value of $\hat{\epsilon}$), and then construct the $T$-spaced sequence given by

$$d(iT) \approx c_0 e^{j2\pi f_o iT + j\theta} s_i + n(iT) \quad (8)$$

## IV. FIRST ORDER CYCLOSTATIONARITY

The estimate of the first-order cyclic mean at the cycle frequency α, $\hat{m}_y^\alpha$ is given by

$$\hat{m}_y^\alpha = \frac{1}{K} \sum_{i=0}^{K-1} y(iT_s) e^{-j2\pi\alpha i} \quad (9)$$

Where $K$ represents the total number of samples used in estimation. Since we only have the sample estimate $\hat{m}_y^\alpha$, we need to pass $\hat{m}_y^\alpha$ through the statistical test proposed in [8] to determine if α is a cycle frequency.

We present the basic steps of the statistical test proposed in [8] to determine if α is a cycle frequency. The presence of a cycle frequency is formulated as a binary hypothesis testing problem: i.e., under hypothesis $H_0$, the tested candidate for cycle frequency is not a cycle frequency, while under hypothesis $H_1$, the tested candidate is a cycle frequency.

- For the candidate cycle frequency α, we estimate $\hat{m}_y^\alpha$ from $K$ samples, and construct the vector

$$\boldsymbol{m}_y = [Re\{\hat{m}_y^\alpha\}, Im\{\hat{m}_y^\alpha\}] \quad (10)$$

- The test statistic

$$\mathcal{T}_K = K \boldsymbol{m}_y \boldsymbol{Q}_y^{-1} \boldsymbol{m}_y^T \quad (11)$$

is then computed. Here, the superscripts $(.)^{-1}$ and $(.)^T$ denote matrix inverse and transpose, respectively, and $\boldsymbol{Q}_y$ is an estimate of the covariance matrix given by

$$\boldsymbol{Q}_y = \begin{bmatrix} Re\left\{\frac{Q_{20}+Q_{21}}{2}\right\} & Im\left\{\frac{Q_{20}-Q_{21}}{2}\right\} \\ Im\left\{\frac{Q_{20}+Q_{21}}{2}\right\} & Re\left\{\frac{Q_{21}-Q_{20}}{2}\right\} \end{bmatrix} \quad (12)$$

Where $Q_{20}$ and $Q_{21}$ are defined as

$$Q_{20} = \frac{1}{L_s K} \sum_{s=-(L_s-1)/2}^{(L_s-1)/2} W(s) F_l\left(\alpha + \frac{s}{K}\right) F_l\left(\alpha - \frac{s}{K}\right)$$

$$Q_{21} = \frac{1}{L_s K} \sum_{s=-(L_s-1)/2}^{(L_s-1)/2} W(s) F_l\left(\alpha + \frac{s}{K}\right) F_l^*\left(\alpha - \frac{s}{K}\right)$$

Where $F_l(f) = \sum_{i=0}^{K-1} y(i+l) y^*(i) e^{-j2\pi f i}$ and $W(s)$ represents a spectral window of length $L_s$.

- Declare α as a cycle frequency, if $\mathcal{T}_K > \Gamma$, where 'Γ' represents a threshold for a given probability of false alarm; else declare α is not a cycle frequency.

## V. FEATURES USED FOR CLASSIFICATION

The digital modulations such as PSK, QAM do not have a first-order cycle frequency since they are generally zero mean. However, AM exhibits first-order cyclostationarity. We use this feature to separate AM from the considered digital modulations.

TABLE I
VALUES OF FEATURES OF INTEREST FOR M-PSK AND M-QAM

| Constellation | $M_{42,s}$ | $|M_{20,s_{dp}}|$ | $|M_{40,s_{dp}}|$ |
|---|---|---|---|
| 2-PSK | 1 | 1 | 1 |
| 4-PSK | 1 | 0 | 1 |
| 8-PSK | 1 | 0 | 0 |
| 16-QAM | 1.32 | 0 | 0.4624 |

The authors in [6] proposed features based on moments to perform classification of M-PSK and M-QAM modulation formats in the presence of frequency offset. In this section, we briefly describe those features. The differentially processed signal is given by

$$s_{dp,i} = s_i s_{i-1}^* \quad (13)$$

The features used in [6] and used in our algorithm are given as follows as

$$M_{21,s} = E[|s_i|^2] \quad (14)$$
$$M_{42,s} = E[|s_i|^4] \quad (15)$$
$$M_{20,s_{dp}} = E[s_{dp}^2] \quad (16)$$
$$M_{40,s_{dp}} = E[s_{dp}^4] \quad (17)$$

The values for the above features for the considered modulation schemes are presented in Table I. The differentially processed sequence of $d(iT)$, defined as $d_{dp}(i)$ is given by

$$d_{dp}(i) = d(i)d^*(i-1) \quad (18)$$

The relationship between the features of $d(iT)$ and the symbols $s_i$ is given by

$$M_{42,s} = \frac{M_{42,d} - 4(M_{21,d} - \sigma^2)\sigma^2 - 2\sigma^4}{(M_{21,d} - \sigma^2)^2} \quad (19)$$

$$|M_{20,s_{dp}}| = \frac{|M_{20,d_{dp}}|}{(M_{21,d} - \sigma^2)^2} \quad (20)$$

$$|M_{40,s_{dp}}| = \frac{|M_{40,d_{dp}}|}{(M_{21,d} - \sigma^2)^4} \quad (21)$$

In practice we only have finite data; the corresponding features are replaced with their sample estimates. For example, the finite data estimate $M_{42,s}$, denoted by $\widehat{M}_{42,s}$ is given by

$$\widehat{M}_{42,s} = \frac{\widehat{M}_{42,d} - 4(\widehat{M}_{21,d} - \sigma^2)\sigma^2 - 2\sigma^4}{(\widehat{M}_{21,d} - \sigma^2)^2} \quad (22)$$

Where $\widehat{M}_{42,d}$ and $\widehat{M}_{21,d}$ are finite data estimates of $M_{42,d}$ and $M_{21,d}$. Similarly we can obtain the estimates of the other features.

**Remark:** We would like to mention here that to fix the threshold, wherever required, independent of the power of the received signal, we normalize it with the square root of the received signal power. The sequence $\{y(iT_s)\}$ represents the normalized signal.

VI. ALGORITHM FOR MODULATION CLASSIFICATION

We now present the steps involved in our algorithm. The algorithm is divided into various stages. We briefly explain the various stages of the algorithm below:

- STAGE-1: In this stage, we evaluate $\widehat{m}_y^\alpha \ \forall \alpha \in [-0.5, 0.5)$, and determine the number of first-order cycle frequencies. If the number of cycle frequencies is 0, we decide in favor of {2-PSK, 4-PSK, 8-PSK, 16-QAM} subclass, else we decide in favor of AM.
- STAGE-2: In this stage, we decide in favor of 2-PSK if $\widehat{M}_{20,s_{dp}} \geq 0.5$, else we decide in favor of {4-PSK, 8-PSK, 16-QAM} subclass.
- STAGE-3: In this stage we separate 16-QAM from {4-PSK, 8-PSK}. If the value of $\widehat{M}_{42,s} \geq 1.155$, then we decide in favor of 16-QAM else we decide in favor of {4-PSK, 8-PSK} subclass.
- STAGE-4: In this stage, we decide in favor of 4-PSK, if $\widehat{M}_{40,s_{dp}} > 0.5$, else we decide in favor of 8-PSK.

VII. ALGORITHM FOR INTELLIGENT RECEIVER

The next stage of classification is the development of an intelligent receiver, which can classify as well as blindly demodulate the received signal. In our system, we have designed an intelligent receiver which can perform classification as well as demodulate AM and 2-PSK signals. We assume *prior* knowledge of the system bandwidth, symbol rate and frame structure in case of the 2-PSK modulation scheme. We now present the basic steps of our algorithm with regard to the intelligent receiver below

- STAGE-1: In this stage, we evaluate $\widehat{m}_y^\alpha \ \forall \alpha \in [-0.5, 0.5)$, and determine the number of first-order cycle frequencies. If the number of cycle frequencies is 0, we decide in favor of 2-PSK, else we decide in favor of AM.
- STAGE-2: In this stage, if the received modulation format was deduced to be AM, we perform a simple envelope detection to extract the information from the received samples, else we pass the received samples through a root-raised cosine filter to perform match filtering and maximize SNR.
- STAGE-3: In this stage, we pass the filtered samples through the Gardner timing recovery loop described in [10] to perform timing synchronization and extract the symbols from the received samples.
- STAGE-4: In this stage we perform fine frequency synchronization using a Costas loop described in [10], to remove the residual frequency and phase offsets from the received samples.
- STAGE-5: In this stage, we perform the SFD detection and the header detection to extract the information from the received samples.

A brief diagram explaining the complete flow of the intelligent receiver is presented in Figure 1.

VIII. SIMULATION RESULTS

*1. Simulation Setup*

We have used 1000 symbols for modulation classification. The transmitted signal $x(t)$ is generated according to Section II. In case of linear modulations such PSK and QAM, $g(t)$ is chosen having a root-raised cosine spectrum with $\rho = 0.5$. The received signal $\{y(iT_s)\}$, is generated as follows

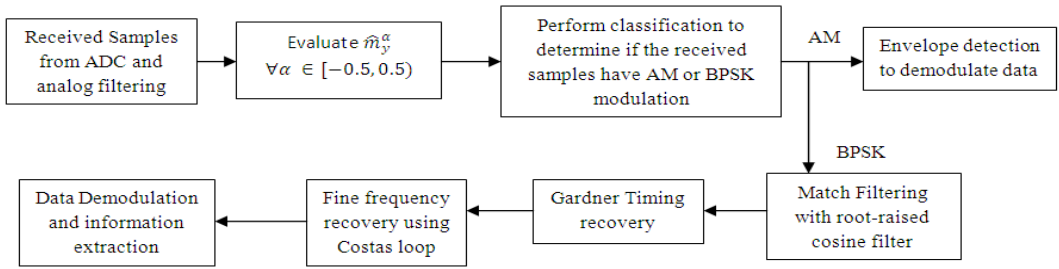

Figure 1: Intelligent receiver design

TABLE II

Percentage of correct classification at an SNR = 5 dB (Number of symbols used = 1000. C1-C5 correspond to 2-PSK, 4-PSK, 8-PSK, 16-QAM, AM)

|    | C1    | C2   | C3  | C4    | C5    |
|----|-------|------|-----|-------|-------|
| C1 | 99.84 | 0    | 0   | 0     | 0.16  |
| C2 | 0     | 100  | 0   | 0     | 0     |
| C3 | 0     | 0    | 100 | 0     | 0     |
| C4 | 44.0  | 1.35 | 0   | 54.65 | 0     |
| C5 | 0.01  | 0    | 0   | 0     | 99.99 |

TABLE III

Percentage of correct classification at an SNR = 10 dB (Number of symbols used = 1000. C1-C5 correspond to 2-PSK, 4-PSK, 8-PSK, 16-QAM, AM)

|    | C1    | C2  | C3  | C4    | C5    |
|----|-------|-----|-----|-------|-------|
| C1 | 99.70 | 0   | 0   | 0     | 0.30  |
| C2 | 0     | 100 | 0   | 0     | 0     |
| C3 | 0     | 0   | 100 | 0     | 0     |
| C4 | 19.84 | 1.9 | 0   | 78.26 | 0     |
| C5 | 0.01  | 0   | 0   | 0     | 99.99 |

$$y(iT_s) = c_0 e^{j2\pi f_o i T_s + j\theta} x(iT_s - \epsilon T) + b(iT_s) \quad (23)$$

Where $c_0$ is a zero mean complex Gaussian random variable of unit variance, $f_o T$ is chosen randomly from the interval [-0.2, 0.2], $\epsilon$ is chosen from the interval [-0.5, 0.5), and θ is chosen from the interval [-π, π]. The value of $c_0$, $f_o T$, and θ are fixed for one realization, but can vary from realization to realization. We computed the sample variance of the first term on the RHS of (23) and generated noise sequence $b(iT_s)$ of appropriate variance to yield a specified SNR. The receive filter $h_{rrc}(.)$ is chosen as a root-raised cosine filter with ρ = 0.5, and of duration 4T. The sequence $\{y(iT_s)\}$ is normalized with the square root of the power in the 1000 symbol-length record. For the statistical test of [8], we chose the spectral window as a rectangular window of length $L_s$ = 61. The threshold Γ used for the detection of the cycle frequency in the test of [8] is set to 15.202, which correspond to a probability of false alarm of $5 \times 10^{-4}$.

*2. Results*

The performance of the classification algorithm is presented in the form of a confusion matrix. Table II gives us the results for an SNR of 5 dB, while Table III gives us the results at an SNR of 10 dB. The performance of the classification algorithm degrades with decreasing SNR, with the degradation in the performance of 16-QAM being the maximum. However, we would like to mention that the performance of the classification algorithm can be improved by increasing the number of symbols considered for classification.

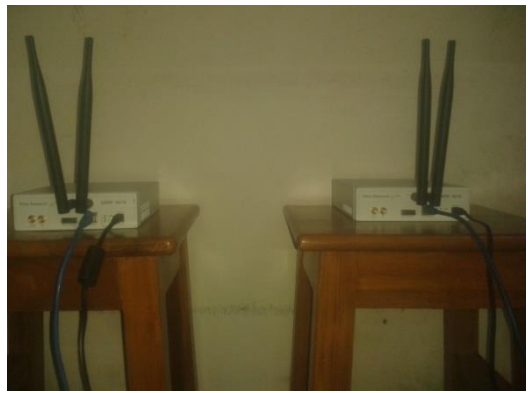

Figure 2: USRP COTS platform Transmitter/Receiver

IX. REAL TIME IMPLEMENTATION IN USRP

The whole system has been implemented on the USRP N210 COTS platform. For the hardware implementation, we assumed a bandwidth of 10 KHz, and a carrier frequency of 400 MHz. We would like to mention here that since the bandwidth of the system is quite small, the noise floor is also significantly less. However, the system performance is limited due to the limited gain in the automatic-gain-control (AGC) before the ADC. Due to the low gain, the total signal-to-quantization-error-ratio (SQNR) of the received samples is very less, which becomes the main bottleneck of the system performance. A snapshot of the USRP N-210 boards in which the system is implemented is presented in Figure 2. Various figures showing our system performing classification and demodulation are shown in Figures 3 and 4.

*1. Modulation Classifier*

For the classification algorithm, no *prior* knowledge of the carrier frequency/phase/timing offsets is assumed. However, we assume *prior* knowledge of symbol rate which is fixed to 9.6 KHz for digital modulation schemes. The classification is done on cluster of samples which correspond to a time interval of 1000 symbols or 104.1 ms. since the statistical test proposed in [8] is computationally expensive; we only pass those values of $\hat{m}_y^\alpha$, whose magnitude exceed a particular threshold, through the test. We fix the value of the threshold as 0.05 in our system. The developed algorithm can also be easily extended to other modulation schemes by increasing the total number of features considered for classification. In order to extend the algorithm for M-FSK schemes, we can use features based on first-order cyclostationarity and to extend it to higher order PSK or QAM modulation formats, we can use either higher order moments or cumulants. But with increasing the number of features used for classification also increases the computational complexity of the classifier.

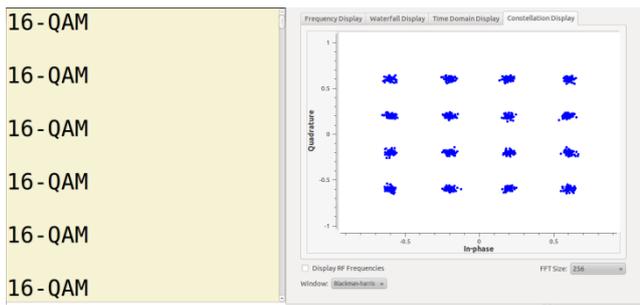

Figure 3: 16-QAM Classification in COTS platform

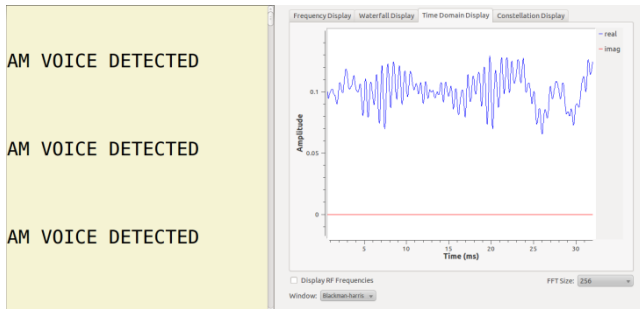

Figure 4: Intelligent receiver demodulating AM/BPSK data

*2. Intelligent Receiver*

For the real time implementation of the cognitive receiver, we assume a symbol rate of 9.6 KHz in the case of 2-PSK. Similar to the classification approach, we only test those cycle frequencies whose magnitude exceeds a particular threshold for the statistical test of [8]. If the received samples are classified as AM, we demodulate them using simple envelope detection, else if the demodulated samples are classified as BPSK. If the received samples are classified as BPSK, we don't perform classification until the complete information from the packet is extracted. The loop bandwidth of the Garner timing recovery loop and the Costas loop are fixed as 0.01 and 0.1 respectively.

## X. CONCLUSIONS

In this paper, we have proposed a new algorithm to perform modulation classification for a 5-class problem consisting of various modulation schemes. The proposed algorithm achieves very good performance for all the considered modulation schemes even at low SNRs and in the presence of various imperfections, highlighting the robustness of the proposed algorithm. The proposed algorithm has also been implemented in a COTS platform and tested in real-time environments. The proposed algorithms are simple to realize in hardware and can perform classification in real-time scenarios.

## BIODATA OF AUTHORS

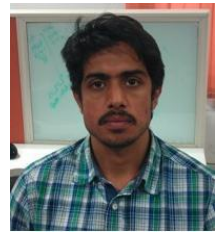

A.K. Karthik received his B.Tech in 2010, M.S (by research) in 2011 from IIIT-Hyderabad, India. He is presently working as a project engineer in UURMI systems Pvt. Ltd. His research interests are in communication system design, detection and estimation theory, blind modulation classification, blind deconvolution and cognitive radios.

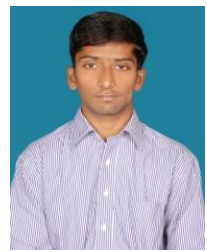

Jameerali M.S received his B.Tech from Jawaharlal Nehru Technological University, India in 2011. He is presently working as a project engineer in UURMI systems Pvt. Ltd. His research interests are in communication system design, detection and estimation theory, blind modulation classification and cognitive radios.

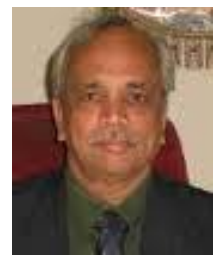

Mr. Bhagavathi Rao is a retired scientist from DLRL, Hyderabad. He has served as director of electronics, R&D headquarters, Delhi and also as the CEO of GATECH, Hyderabad, and group head of electronic warfare systems at DLRL among other assignments. He is presently associated with Uurmi Systems guiding the advanced development of next generation communication and radar systems.